\documentclass[twocolumn,showpacs,preprintnumbers,amsmath,amssymb,aps,pre,superscriptaddress]{revtex4-2}

\usepackage{graphicx}
\usepackage{dcolumn}
\usepackage{bm}
\usepackage{tikz}
\usepackage{bbm}
\usepackage{ulem}

\usepackage{comment}

\begin{document}


\title{Multiscale Relevance of Natural Images}

\author{Samy Lakhal}
   \affiliation{Chair of Econophysics \& Complex Systems,  Ecole Polytechnique, 91128 Palaiseau Cedex, France}
 \affiliation{LadHyX, {UMR} CNRS 7646, Ecole Polytechnique, 91128 Palaiseau Cedex, France}
  \affiliation{Institut Jean Le Rond d’Alembert, UMR CNRS 7190, Sorbonne Université, 75005 Paris, France}

\author{Alexandre Darmon}
 \affiliation{Art in Research, 33 rue Censier, 75005 Paris, France}

\author{Iacopo Mastromatteo}
\affiliation{Chair of Econophysics \& Complex Systems,  Ecole Polytechnique, 91128 Palaiseau Cedex, France}
\affiliation{Capital Fund Management, 23 rue de l'Université, 75007 Paris, France\smallskip}

\author{Matteo Marsili}
\affiliation{Quantitative Life Sciences Section\\
The Abdus Salam International Centre for Theoretical Physics, 34151 Trieste, Italy\medskip}

\author{Michael Benzaquen}
 \email{michael.benzaquen@polytechnique.edu}
 \affiliation{Chair of Econophysics \& Complex Systems,  Ecole Polytechnique, 91128 Palaiseau Cedex, France} \affiliation{LadHyX, {UMR} CNRS 7646, Ecole Polytechnique, 91128 Palaiseau Cedex, France}
 
 \affiliation{Capital Fund Management, 23 rue de l'Université, 75007 Paris, France\smallskip}

\date{\today}

\begin{abstract}
We use an agnostic information-theoretic approach to investigate the statistical properties of natural images. 
 We introduce the Multiscale Relevance (MSR) measure~\cite{cubero_multiscale_2020} to assess the robustness of images to compression at all scales. 
Starting in a controlled environment, we characterize the MSR of synthetic random textures as function of image roughness $\text H$ and other relevant parameters. We then extend the analysis to natural images and find striking similarities with critical ($\text{H}\approx 0$) random textures. 
We show that the MSR is more robust and informative of image content than classical methods such as power spectrum analysis.
Finally, we confront the MSR to classical measures  for the calibration of common procedures such as color mapping and denoising.
Overall, the MSR approach appears to be a good candidate for advanced image analysis and image processing, while providing a good level of physical interpretability.
\end{abstract}

\maketitle

\tableofcontents

\section*{Introduction}

Recent advances in image processing have benefited from the emergence of powerful learning frameworks combining efficient architectures~\cite{krizhevsky_imagenet_2017,lecun_gradient-based_1998,vaswani_attention_2017} with large high-quality databases~\cite{krizhevsky_learning_2009,deng_imagenet_2009}. In particular, neural networks, layering  simple linear and non-linear operators such as convolution matrices or activation functions, have proven to be very efficient to classify or generate high dimensional data. They are now able to capture similarities between images with unprecedented success. However, while their performance increases with the depth of the architecture, it is generally at the cost of physical interpretation. Understanding the learning dynamics and the statistical features of the resulting images remains a challenge for the community \cite{saxe_exact_2014,goldt_dynamics_2019}.

Before the advent of machine learning algorithms, tasks such as compression~\cite{wallace_jpeg_1992,skodras_jpeg_2001}, denoising~\cite{rudin_nonlinear_1992} or edge detection were (and in some cases still are) performed using signal processing methods. Among the classical approaches, the first kind is based on specific measures, such as the widely used  Peak Signal-to-Noise Ratio (PSNR)~\cite{korhonen2012peak}, that are built upon common signal processing metrics (Euclidian distance, power spectrum, etc.). The second family uses vision based experiments to construct semi-empirical measures of similarities, such as the Structural Similarity Index (SSI)~\cite{wang_image_2004}. In both cases the approach is fully deterministic, which means that stochastic properties like roughness, stationarity, or local correlations are ignored.

In the context of statistical physics, the problem of high dimensional data inference has recently been addressed using a novel, fully agnostic, approach. Developed to measure specific properties of finite size samples~\cite{marsili_quantifying_2022},  the approach  consists in assessing the influence of a prescribed compression procedure over simple entropy measures. Applications in biological inference~\cite{cubero_multiscale_2020}, finance~\cite{haimovici_criticality_2015}, language models~\cite{marsili_quantifying_2022} or optimal machine learning~\cite{song_resolution_2018,duranthon_maximal_2021} have already shown exciting results.  In this paper, we adapt the latter formalism to image analysis and image processing, focusing specifically on the case of natural images. Natural scenes or landscapes have long been studied as they display distinguishable statistical features such as scale invariance~\cite{balboa_power_2003,ruderman_statistics_1994,ruderman_statistics2_1994}, non-Gaussianity~\cite{zoran_scale_2009}, or patch criticality~\cite{stephens_statistical_2013}.

The outline of the paper is as follows. In Section~\ref{sec:ResolutionRelevanceFramework}, we  introduce the Resolution/Relevance formalism using an illustrative example, and adapt it to the purpose of image analysis.  In Section~\ref{sec:1/f}, we analyse a class of parameterizable images, that is random $1/f^\alpha$ Gaussian fields, and introduce the Multiscale Relevance (MSR). In Section~\ref{sec:natural}, we extend the analysis to natural images and their gradient magnitudes. We discuss meaningful statistical similarities with the synthetic Gaussian fields. In Section~\ref{sec:ImageProcessing}, we show how the MSR approach can be used in the context of common image processing tasks.

\section{The Resolution/Relevance framework }
\label{sec:ResolutionRelevanceFramework}

Here we present the information-theoretic framework that was recently built by one us~\cite{marsili_quantifying_2022} for agnostic analysis of high-dimensional data samples and their behaviour under compression procedures. Relevant metrics are derived from simple statistics of the compressed samples.

\subsection{Tradeoff between precision and interpretability}

Let us consider the  problem of binning, namely clustering samples of a random variable $X$ into groups characterized by a similar value of $X$. If the sampled data points $\mathcal S= \{x_1, \dots, x_N\}$ all take different states (e.g. when the distribution of $X$ is continuous) the empirical distribution is a Dirac comb. In order to gain insight into the sampled variable, one can visualize the data by using histograms with well chosen bins/boxes. Indeed, this procedure enforces the emergence of structure by reducing data resolution through compression, allowing for more interpretability. One can then make assumptions on the underlying process and find the optimal parameters to best describe the data.

\begin{figure}
    \centering
    \includegraphics[width = \linewidth]{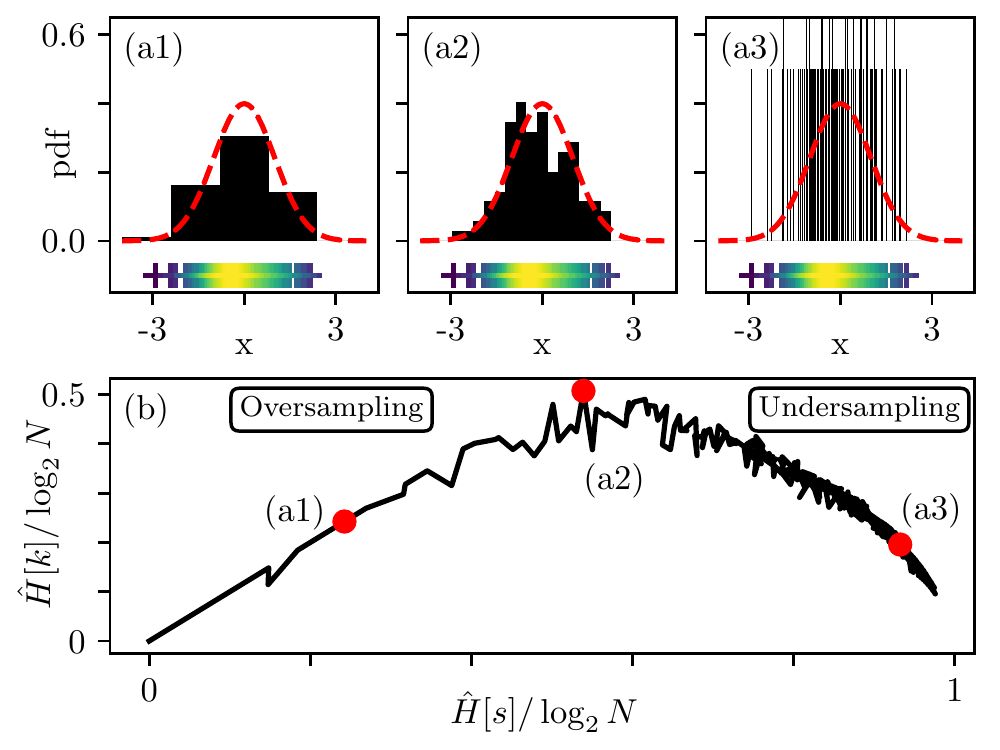}
    \caption{Relevance analysis of a Gaussian distribution sample ($N= 100$). (a) Influence of the number of bins $n$ on the normalized histogram (black bars), for (a1) $n=5$, (a2) $n=23$ and (a3) $n=400$. The red curve corresponds to the underlying distribution. The bottom markers ($+$) represent the initial sample data points with color indicating local data density. (b) Resolution/Relevance curve.}
    \label{fig:HsHkGaussian}
\end{figure}

We illustrate this intuition by sampling $N=100$ realizations of a Gaussian variable $X\sim\mathcal{N}(0,1)$ in Fig.~\ref{fig:HsHkGaussian}. The data are binned into $n$ identical boxes, for three different values of $n=5$, $23$ and $400$. We also define the bin width  $\ell$ as a compression parameter transforming the original sample $\mathcal S$ into a compressed sample $\mathcal S^{\ell}$. The compression step consists in replacing each data point by its corresponding histogram bar index.
Figure~\ref{fig:HsHkGaussian}(a1) (large $\ell$) displays a situation of \textit{oversampling}. With only five bins a considerable amount of data resolution is lost. 
On the contrary, Fig.~\ref{fig:HsHkGaussian}(a3) (small $\ell$) corresponds to an \textit{undersampling} regime, with very narrow bins (mostly containing only one data point) and a resulting distribution close to a Dirac comb. 
Figure~\ref{fig:HsHkGaussian}(a2) (intermediate $\ell$) appears as a reasonable compromise in which the histogram is visually close to the generator density, indicating we might be close to the optimal level of data compression.  From the latter observation, one is tempted to go for a Gaussian model, with suitable estimators for the mean and variance. However such decision  solely relies on a specific  compression level, and thus does not make full use of the sample at play.

The formalism that we  introduce in the next section  provides a principled framework to connect the choice of the compression level with an optimality criterion that is agnostic to the nature of the generative model from which the data is sampled.

\subsection{Resolution and Relevance}

Previous work from Marsili \textit{et al.} \cite{haimovici_criticality_2015} addressed the issue of the  \textit{overampling/undersampling} transition by introducing observables that  allow one to monitor changes in a reduced sample $\mathcal{S}^{\ell}=\{s^{\ell}_1,\dots ,s^{\ell}_N\}$ obtained by compressing $\mathcal{S}$ with a parameter $\ell$. First, let us consider $k^{\ell}_s$ the number of data points of identical state $s$ and $m^{\ell}_k$ the number of states appearing $k$ times in $\mathcal{S}^{\ell}$. 
It follows that $\sum_s k^{\ell}_s = \sum_k k m_k^{\ell} = N$. For example, in the compressed sample displayed in Fig.~\ref{fig:HsHkGaussian}(a1), values taken by $k^{\ell}_s$ are $\{2,26,49,23,0\}$, and since each bar in the histogram has a different height, one has $m_2=m_{26}=m_{49}=m_{23}=m_{0} = 1$ and $m_k=0$ otherwise.

One can then define the \textit{Resolution} $\hat H^{\ell}[s]$ and \textit{Relevance} $\hat H^{\ell}[k]$ as:
\begin{equation}
    \left\{
    \begin{array}{lll}
    \hat H^\ell[s] & = - \sum_s \frac{k^\ell_s}{N} \log_2 \frac{k^\ell_s}{N}, \\
    
     \hat H^\ell[k] & = - \sum_k \frac{k m^\ell_k}{N} \log_2 \frac{k m^\ell_k}{N}.
    \end{array}
\right.
\label{eq:HsHk}
\end{equation}
The \textit{Resolution} is the entropy of the empirical distribution $\{p_s^{\ell} = k_s^{\ell}/N\}_s$  and describes the average amount of bits needed to code a state probability in $\mathcal{S}^{\ell}$. The compression clusters data points together hence reducing the average coding cost. The \textit{Resolution} is maximal for raw data and monotonically decreases with $\ell$, until it reaches the minimally entropic fully compressed sample. The \textit{Relevance} is the entropy of the distribution $\{q^{\ell}_k = k m^{\ell}_k/N\}_k$, that is the probability that a data point sampled from $\mathcal{S}^{\ell}$ appears $k$ times in the sample. This is a compressed version of $p^{\ell}_s$, where identical frequency states are clustered, dropping their label $s$ in the process. Knowing $q^{\ell}_k$ is then sufficient to build a histogram without labels, and is equivalent to assuming indistinguishability of states sampled the same number of times. Sorting them in decreasing frequency values would yield the famous Zipf plot. In the end, the \textit{Relevance} encodes the height of each bar and is maximal when $\{km^{\ell}_k/N\}_k$ is uniformly distributed, leading to $m_k \propto k^{-1}$. 
We reported in Tab.~\ref{tab:OversamplingUndersampling} the typical sampling situations and their corresponding value in Resolution/Relevance.

\begin{table}[b!]
\caption{\label{tab:OversamplingUndersampling}%
Typical sampling situations.}
\begin{ruledtabular}
\begin{tabular}{ccccc}

Situation&Sampled States&$\{m_k\}_k$&$H[s]$& $H[k]$\\
\hline
\begin{tabular}{@{}c@{}}Full \\ Oversampling\end{tabular} & Identical &$\left.
    \begin{array}{ll}
    m_N =1\\
    m_k =0\end{array}\right.$&0&$0$\\ \hline
\begin{tabular}{@{}c@{}}Full \\ Undersampling\end{tabular} & Distinct & $\left.
    \begin{array}{l}
    m_1  =N\\
    m_k = 0\end{array}\right.$ & $\log N$& $0$ \\ \hline 

 \begin{tabular}{@{}c@{}}Intermediate \\ sampling  \end{tabular}  & Intermediate  & $m_k \propto k^{-1}$ & $H_0$ & $\max H[k]$ \\

\end{tabular}
\end{ruledtabular}
\end{table}

Coming back to the Gaussian sampling example,  Fig.~\ref{fig:HsHkGaussian}(b) displays $\hat H^{\ell}[k]$ as function of $\hat H^{\ell}[s]$, obtained by varying $\ell$. Corresponding values for $n=5$ (a1), $n=23$ (a2) and $n=400$ (a3) are highlighted. Note that (a2) maximizes Relevance while (a1) and (a3) respectively correspond to oversampling and undersampling. Let us emphasize at this point, that, despite the visual impression in this specific example,  the sample (a2) does not necessarily minimize the distance between the underlying and empirical distributions. Interestingly, the Resolution/Relevance properties are only dependent on the raw sample $\mathcal{S}$ and the compression parameter $\ell$,  making the overall approach agnostic to the generating process.  What is most interesting is thus the way in which the sample evolves with compression, while transitioning  from {undersampling} to {oversampling}. As a result, one must choose a compression procedure that allows to crossover between these two regimes.

\subsection{Application to images}

 Images are usually described as  fields $h(\bm r)$ where $\bm r \in \{1, \dots, N_X\}\times\{1,\dots, N_Y\}$. This is equivalent to a sample made of $\mathcal{S} = \{ (\bm r,h(\bm r)\} $ of size $N = N_X N_Y$, describing the position and color of each pixel. Naturally, $\mathcal{S}$ lies in the full undersampling regime as each data point is unique.

\begin{figure}
    \centering
    \includegraphics[width = \linewidth]{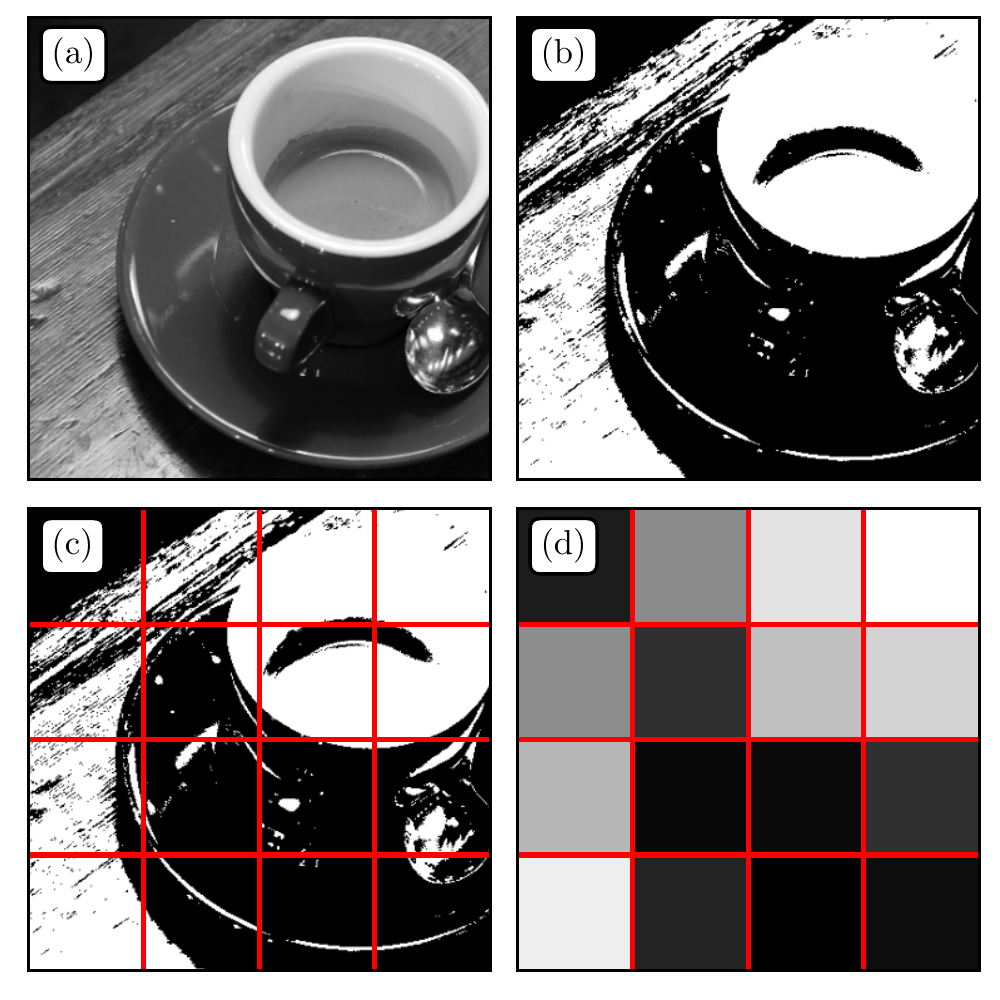}
    \caption{Illustration of the segmentation/compression procedure on a classic benchmark image. (a) Original Image. (b) Thresholded image at a given quantile value $a$. (c) Thresholded image with reduced grid. (d)  Reduced sample  where each grid cell is replaced by the average pixel color.}
    \label{fig:IllustrationCompression}
\end{figure}

To compress grayscale images, we therefore propose a simple procedure consisting in two steps: (i) segmentation, and (ii) spatial compression, as illustrated in Fig.~\ref{fig:IllustrationCompression}. Segmentation means grayscale levels are transformed into black and white pixels using a threshold level $a$ (fraction of black pixels), leading to the binary image $h_a(\bm r)$ (Fig.~\ref{fig:IllustrationCompression}(b)). This lowers the amount of possible color states in the sample, a necessary condition to reach the full oversampling regime. Note that one can reconstruct the original image by averaging over all segmentations. This step is generalisable for colors, for example by using a triplet $(a_{\mathrm R},a_{\mathrm G},a_{\mathrm B})$ in the RGB space. 
The second step consists in the compression of pixel positions (Fig.~\ref{fig:IllustrationCompression}(c)). One replaces each coordinate $\bm r$ by the index $\bm r^\ell$ of its position on a grid of stepsize $\ell$. One ends up with a compressed sample:
\begin{equation}
    \mathcal{S}^{\ell}_a = \{(\bm r^\ell,h_a(\bm r))\}.
    \label{eq:SampleCG}
\end{equation}
Each pixel value is then replaced by its average in the reduced grid  (Fig.~\ref{fig:IllustrationCompression}(d)). 
Finally, $k^\ell_{(\bm r_\ell,0)}$ and $k^\ell_{(\bm r_\ell,255)}$ would be defined as the number of black and white pixels in cell $\bm r^\ell$, and $m_k^\ell$ as the number of cells with $k$ black or white pixels at scale $\ell$. Using Eqs~\eqref{eq:HsHk}, one can compute the values of $\hat H^{\ell}[s]$ and  $\hat H^{\ell}[k]$ that will be used in the sequel.

One can make a direct analogy between this compression procedure and image processing architectures such as Convolution Neural Networks (CNN) \cite{krizhevsky_imagenet_2017}. First, their constitutive layers usually combine a spatial compression procedure, that is a first linear convolution, with a trainable or prescribed layer. Then, a segmentation step is performed using a nonlinear transformation on pixel values called \textit{activation function}. In a similar fashion, our procedure consists in a one layer network, taking $\mathcal{S}$ as input and giving $\mathcal{S}^\ell_a$. Interestingly, we do not need to specify a particular convolution matrix as an input to the algorithm, but only a size parameter, by that making our approach more agnostic.
Ultimately, note that any compression procedure allowing the undersampling/oversampling transition could have been selected. For example, one could use Discrete Fourier or Wavelet coefficients, classically used in JPEG compression algorithms~\cite{skodras_jpeg_2001,wallace_jpeg_1992}. Another approach would consist in using  intermediate representations of trained or untrained networks with binary activation functions (perceptron-like) and tunable layer size, as in the Resolution/Relevance trade-offs of deep neural architectures~\cite{song_resolution_2018}.

\section{Relevance of random textures \label{sec:1/f}}

In this section we illustrate the use of the metrics ($\hat H^{\ell}[s]$, $\hat H^{\ell}[k]$) on a simple yet widely encountered class of processes: two-dimensional $1/f^\alpha$ random Gaussian fields. We first recall the properties of such fields and then study the influence of $\alpha$ on Resolution and Relevance.

\subsection{On $1/f^\alpha$ Gaussian fields}
\begin{figure}[t!]
    \centering
    \includegraphics[width = \linewidth]{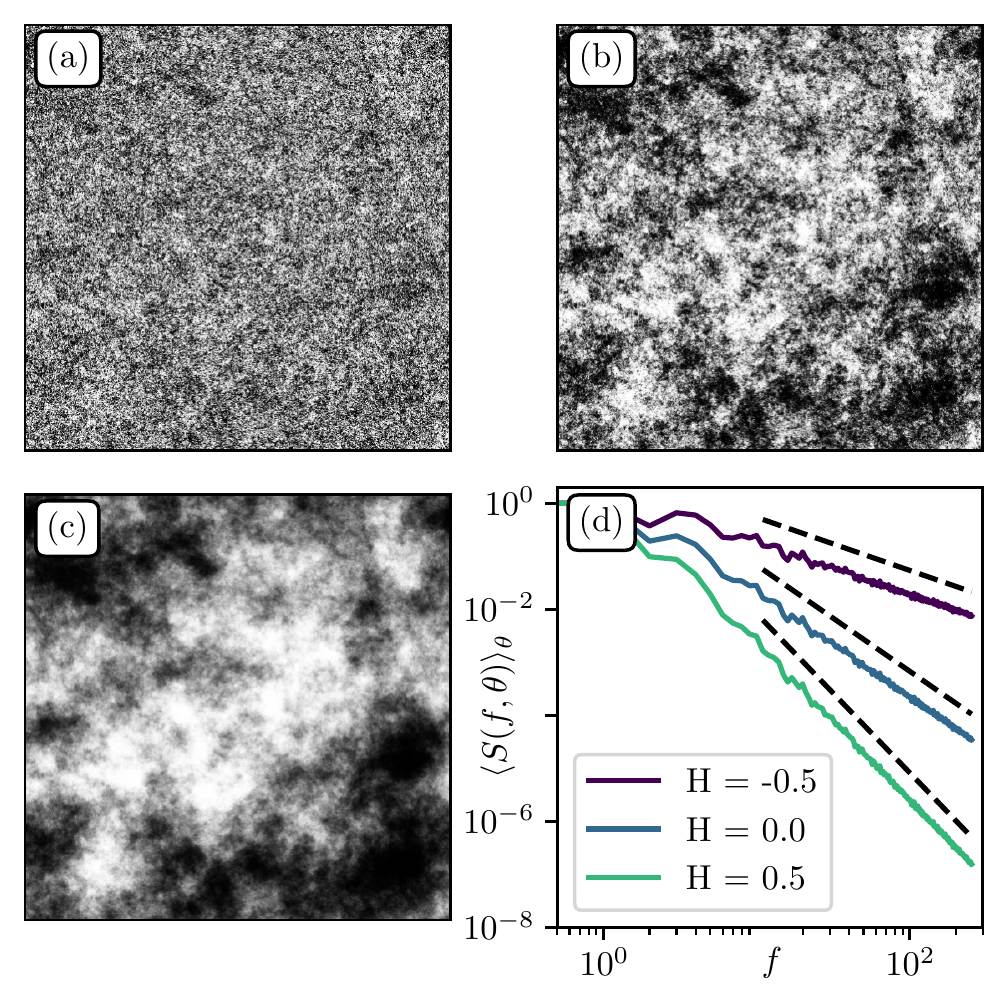}
    \caption{$1/f^\alpha$ textures generated from the same white Gaussian noise seed. (a), (b), (c) Representations of  $1/f^\alpha$ random fields with respective roughness $\text{H} = -0.5, 0, 0.5 $ and spatial resolution $512\times512$. (d) Azimuthally averaged power spectrum $\langle S(f,\theta) \rangle_\theta$. Black dashed lines indicate the theoretical power spectrum decay $1/f^{2 \alpha}$ with  $\alpha = 1+\text{H}$.}
    \label{fig:powerlaw}
\end{figure}

  $1/f^\alpha$ Gaussian fields consist in the linear filtering of an initially uncorrelated 2D white noise (see  Appendix.~\ref{appendix:1/f}). The latter presents a flat Fourier spectrum that is then multiplied by $1/f^\alpha$, therefore leading to a power spectrum scaling as $1/f^{2\alpha}$. This leads to the forcing of spatial correlations in the direct space.  Such power law filter introduces scaling properties that are usually described by the roughness \textit{Hurst} exponent $\text{H} := \alpha - d/2$ where $d$ is the field dimension (here $d=2$). Depending on the sign of $\text{H}$, one can recover two types of processes. When $\text{H}<0$ the random field is stationary, that is with fixed mean and correlations $C(\delta r) \propto \delta r^{2\text{H}}$ at lag distance $\delta r$. The specific case $\text{H} = -d/2$ corresponds to an unmodified spectrum (white noise). When $\text{H}>0$, the process is no longer stationary but possesses stationary increments with scaling $\langle [ h(\bm r+ \delta \bm r)-h(\bm r )]^2\rangle \propto \delta r^{2\text{H}}$. We generate three samples of distinct roughness values $\text{H}\in \{-1/2,0,1/2\}$, shown in Fig.~\ref{fig:powerlaw}(a), (b) and (c) respectively. The Hurst exponent influences the visual aspect of roughness, with images getting smoother as H increases. Figure~\ref{fig:powerlaw}(d) shows the azimuthally averaged power spectrum $S(f) = \langle |\tilde h(f,\theta)|^2\rangle_{\theta}$ allowing to check that the generating method is robust as the expected scaling behavior and exponents are recovered.  

\subsection{Multiscale Relevance of random textures}

\begin{figure}[h]
    \centering
    \includegraphics[width = \linewidth]{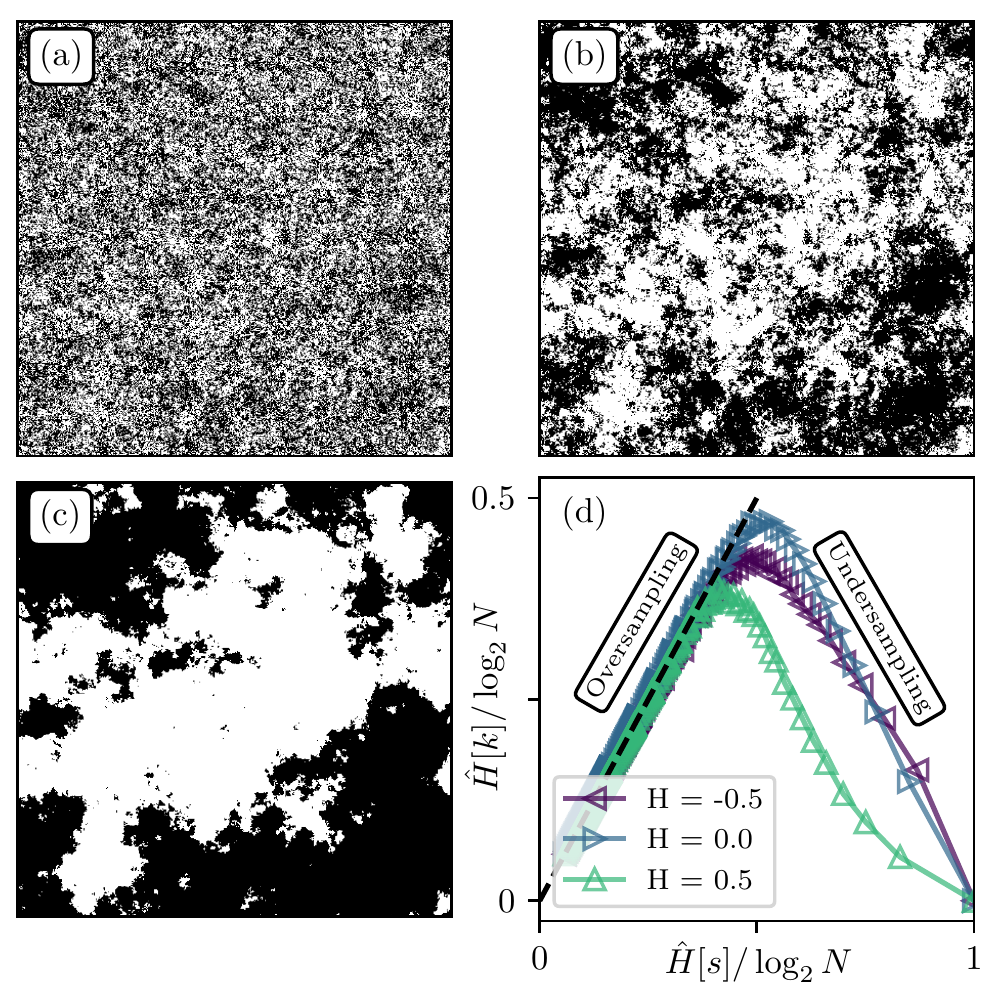}
    \caption{(a), (b), (c) Segmented versions of the textures of Fig.~\ref{fig:powerlaw}, with $\text{H} = -0.5, 0, 0.5 $ respectively, and threshold value $a=0.5$. (d) Resolution/Relevance curves normalized by the maximum entropy $\log_2 N$.}
    \label{fig:powerlaw_msr}
\end{figure}

We now perform the segmentation described above on the fields presented in Fig.~\ref{fig:powerlaw}. The resulting textures for threshold value $a=0.5$ are displayed in Fig.~\ref{fig:powerlaw_msr}(a)-(c) and the corresponding Resolution/Relevance curves $(\hat H^\ell[s],\hat H^\ell[k])_{\ell \in \{1,\dots, N\}}$ are plotted in Fig.~\ref{fig:powerlaw_msr}(d).  

One can see that while the patterns remain quasi-identical for $\text{H} = -0.5$ (Fig.~\ref{fig:powerlaw_msr}(a)) and $\text{H}=0$ (Fig.~\ref{fig:powerlaw_msr}(b)), this is not the case for $\text{H}=0.5$ (Fig.~\ref{fig:powerlaw_msr}(c)) where large areas of uniform tint are created by the segmentation procedure. This is due to the presence of stronger spatial correlations, inducing more persistence of patterns and less fluctuations around the average. Further, one can see that the $\text{H}=0$ texture displays interesting visual features at all scales, as reported in visual quality assessment experiments \cite{lakhal_beauty_2020},  while they appear limited to small scales for $\text{H}=-0.5$. It is not straightforward to connect these observations with the Relevance curves in  Fig.~\ref{fig:powerlaw_msr}(d), as the relative Relevance varies with Resolution. It thus seems more natural to consider the Relevance across all levels of compression.

To do so, we introduce a measure that quantifies the overall robustness of a sample to compression called \textit{Multiscale Relevance} (MSR) and defined as:
\begin{equation}
    \text{MSR} := \int \hat H^{\ell}[k] d\hat H^{\ell}[s],
    \label{eq:MSR}
\end{equation}
which is non other than the area under the Resolution/Relevance curve. 
This measure was introduced in  \cite{cubero_multiscale_2020} as an order parameter characterizing neuronal activity time series, and was successful at distinguishing useful information from ambient noise, as expected from a complexity measure~\cite{aaronson_quantifying_2014}.
Note that while several measures of complexity based on multi-scale entropy contributions have already been introduced in the literature \cite{zhang_complexity_1991,humeau-heurtier_multiscale_2020}, the MSR differs in that the contribution of each scale is naturally weighted by the Resolution. Other measures generally give identical weights to each compression level. 
\begin{figure}[t!]
    \centering
    \includegraphics[width = \linewidth]{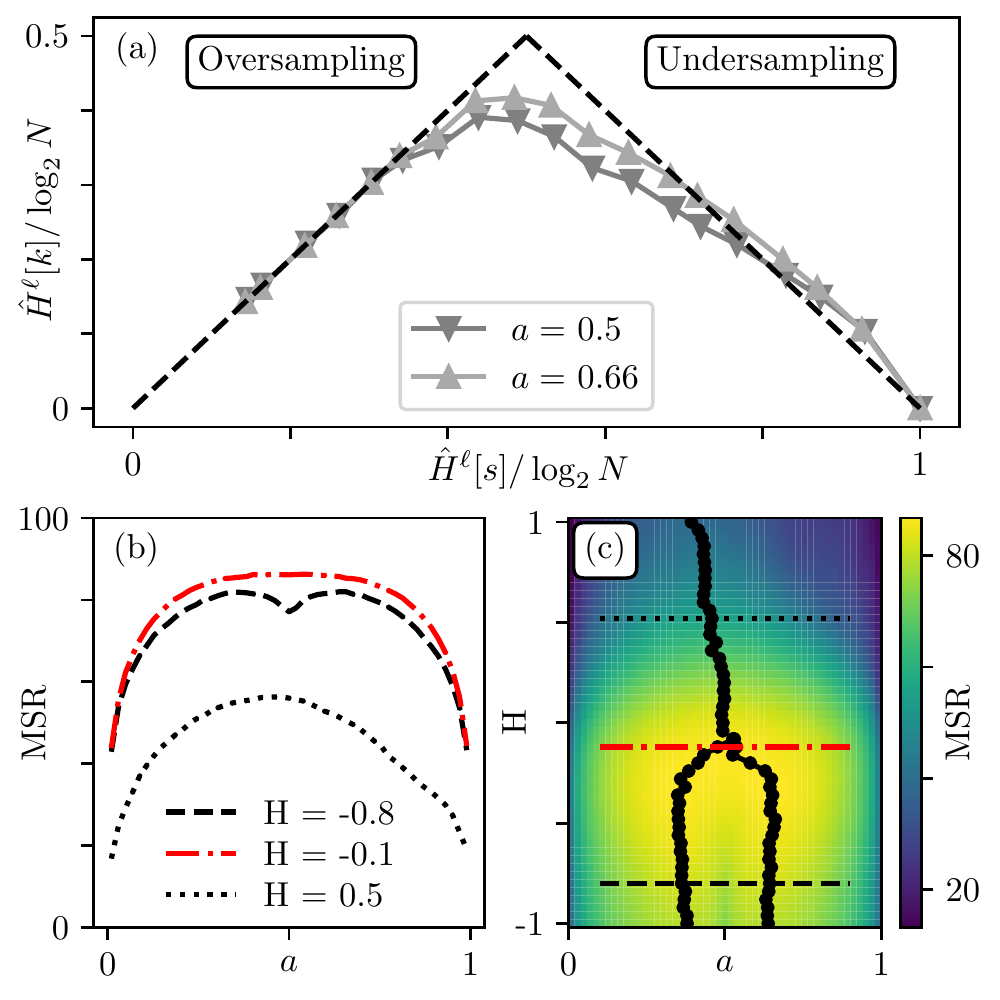}
    \caption{Influence of the segmentation value $a$. (a) Relevance curves for $\text{H}=-0.8$  for two values of $a$. (b)  MSR as function of $a$ for $\text{H}=-0.8$ (black dashed line), $\text{H}=-0.1$ (red dashed dotted line) and $\text{H}=0.5$ (black dotted line). (c) Density plot MSR$(\text{H},a)$. The maxima are signified with black markers.}
    \label{fig:percolation}
\end{figure}
For the images in Fig.~\ref{fig:powerlaw_msr}, one obtains $\text{MSR}(\text{H}=0.5)<\text{MSR}(\text{H}=-0.5)<\text{MSR}(\text{H}=0)$. This is consistent with our previous visual impression that the texture in Fig.~\ref{fig:powerlaw}(b) seems to contain more information at different scales.

\subsection{Most relevant segmentation(s)}

One naturally expects the segmentation threshold $a$ to influence the Relevance. Indeed, at given $\text{H}<0$, most relevant representations do not seem to correspond to $a=0.5$. This is confirmed in Fig.~\ref{fig:percolation}(a) where the Relevance curve for $H=-0.8$ is higher for $a=0.66$ than $a=0.5$. 
Figure~\ref{fig:percolation}(b) displays the MSR as function of $a$ for three values of H. For $\text{H}=-0.8$ (dashed curve) one observes two symmetric maxima at $a_c = 0.5 \pm .13$, consistent with Fig.~\ref{fig:percolation}(a). Interestingly,  breaking the symmetry in the distribution of pixels by choosing a ``background canvas" leads to more interesting samples in terms of Resolution/Relevance. As one can see in Fig.~\ref{fig:percolation}(c), there is a bifurcation at $\text{H}\approx 0$ below which two maxima of MSR coexist. The obtained values of $a_c$ for $\text{H}<-1/2$ fall close to the classic percolation threshold $a^* = 0.59$ on the 2D square lattice \cite{isichenko_percolation_1992}. Indeed, our segmented images are equivalent to samples of the correlated percolation site problem. In particular, Prakash~\textit{et al.}~\cite{prakash_structural_1992} observed, as we do here, that when $\text{H}\to 0$ from below both maxima continuously meet at $a_c = 0.5$ while flattening the MSR$(a)$ curve around such value (see Fig.~\ref{fig:percolation}(b)). At this critical point, the information content of images becomes less sensitive to the segmentation process. 

When $\text{H}\gtrsim 0$, MSR$(a)$ displays one unique maximum at $a_c=0.5$. However, as $\text{H}$ increases further, so does the range of correlations,  leading to finite-size effects. The resulting $a_c$ becomes very noise dependent as different samples lead to different critical thresholds. Interestingly, such behavior was also reported in the percolation of 2D Fractional Brownian Motion~\cite{du_percolation_1996}.

\section{Relevance of natural images\label{sec:natural}}

We now focus on \textit{natural images}, namely pictures of natural scenes and landscapes. These have long been studied in the literature~\cite{ruderman_statistics_1994,ruderman_statistics2_1994,zoran_scale_2009,van_der_schaaf_modelling_1996,balboa_power_2003}, as they display robust statistical features, such as scale invariance and criticality.

\begin{figure}[b!]
    \centering
    \includegraphics[width = \linewidth]{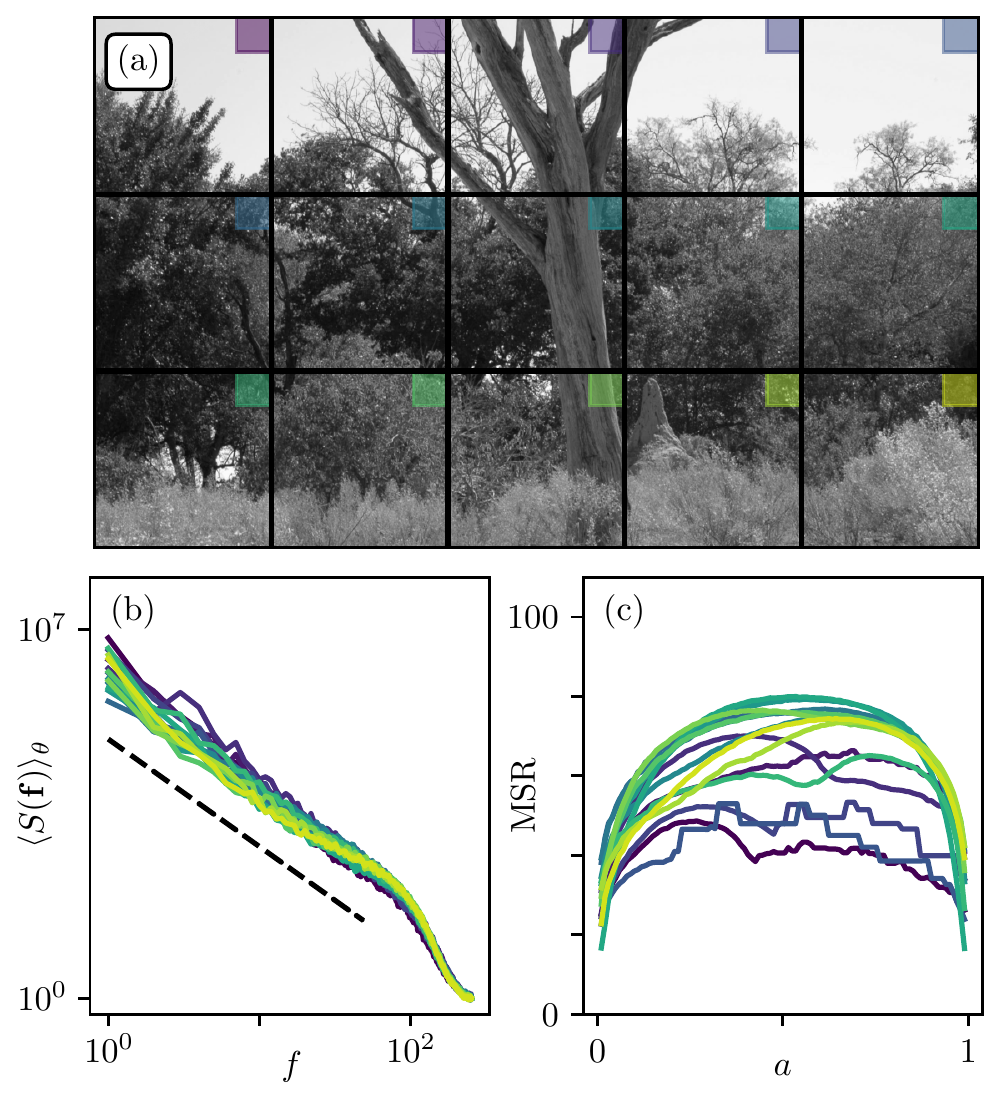}
    \caption{(a) Natural grayscale image from \cite{tkacik_natural_2011}, segmented in patches of size 512$\times$ 512. (b) Power spectrum for each patch. Dotted line is a decaying power law with exponent $- 2$. (c)~MSR as function of $a$ for each patch.}
    \label{fig:natural}
\end{figure}

\subsection{On the grayscale field $h(\bm r)$}
Figure~\ref{fig:natural}(a) shows the photograph from Tkacik~\textit{et al.}~\cite{tkacik_natural_2011} in the Okavango Delta in Botswana, described as a ``[...] \textit{tropical savanna habitat similar to where the human eye is thought to have evolved}".  The image is subdivided into fifteen patches of size $512\times512$ pixels. One can observe a wide variety of patterns, ranging from uniform shades of light gray in the sky to strong discontinuities with tree branches and noisy vegetation textures.

A power spectrum analysis for all patches is shown in Fig.~\ref{fig:natural}(b). The shape in the high frequency limit is due to camera calibration, optical blurring, or post-processing procedures, which are  independent of the patch content. At low frequency we  observe a decaying power law with exponent $-2.0 \pm 0.1$. Note that, although there are small fluctuations that may be related to patch features  \cite{van_der_schaaf_modelling_1996}, the power spectrum analysis seems rather unable to capture the visual heterogeneity from one patch to another mentioned above. 

 This being said, $S(f) \sim 1/f^{2}$ translates to $\text{H} = 0.0 \pm 0.1$ in terms of roughness exponents, which is precisely the range in which the MSR displayed critical and nontrivial behaviour for random textures in Sec.~\ref{sec:1/f}.
\begin{figure}[b!]
    \centering
    \includegraphics[width = \linewidth]{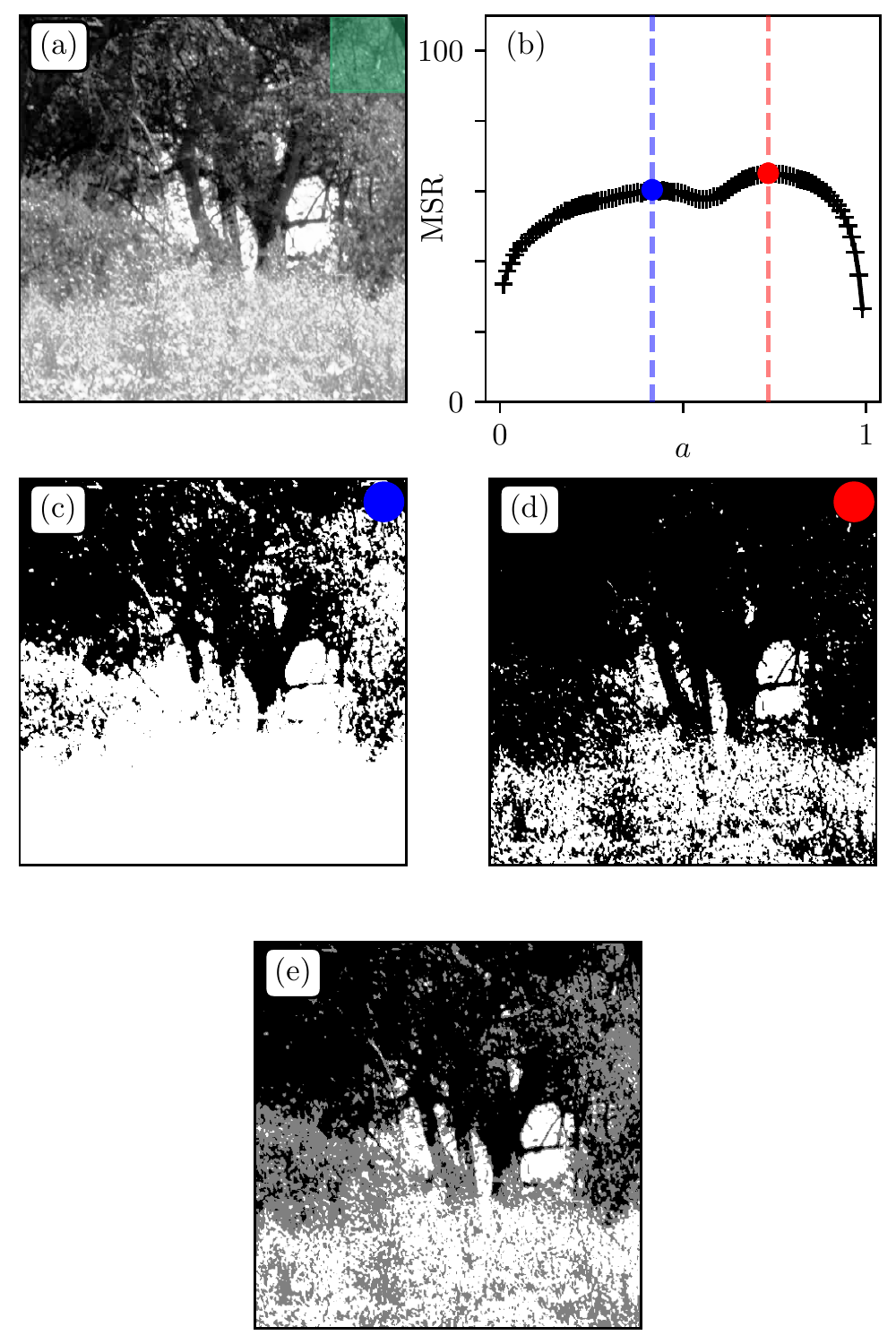}
    \caption{(a) Bottom-left patch of Fig.~\ref{fig:natural}(a). (b) MSR as function of $a$ with highlighted critical thresholds $(a_1,a_2)=(0.42,0.73)$. (c)(d) Corresponding segmented patches. (e) Image obtained by adding  (c) and (d), with  three color levels \{0,127,255\}.}
    \label{fig:Segmentation}
\end{figure}
We thus expect that the MSR approach may allow for a finer characterization of each patch.  Another issue with classical spectral analysis is that the power spectrum of the image is expected to be extremely sensitive to non-linear transformations of its color histogram, even monotonous, that keep the visuals identical. With the MSR method, there is no such issue  as the segmentation parameter $a$ defines the proportion of black and white pixels, regardless of the shape of the color histogram.

Figure~\ref{fig:natural}(c) shows the MSR curves for all patches. First observation is that the range of MSR values is similar in magnitude to that of $\text{H}\approx 0$ textures in Sec.~\ref{sec:1/f}. Then, one clearly sees significant differences between the MSRs of each patch. Patches containing mainly bushy textures with no abrupt changes in patterns display a unique maximum in the MSR($a$) curve. Note that the singularities that appear in some cases are due to specific colors being disproportionate in the histogram (uniform sky). Patches containing heterogeneous shades, or physical objects of different sizes combining tree truncs, branches and bush (e.g. bottom left in \ref{fig:natural}(a)) tend to display two maxima, similarly to $\text{H}<0$ (see  Sec.~\ref{sec:1/f}).

Figure~\ref{fig:Segmentation}  focuses on the bottom-left patch of Fig.~\ref{fig:natural}(a). This sub-image seems to display two distinct dominant color levels. Such levels actually correspond to the maxima of the MSR curve in Fig.~\ref{fig:Segmentation}(b). This is visually confirmed from the segmentations \ref{fig:Segmentation}(c) and \ref{fig:Segmentation}(d) which capture best the fluctuations at the top  and bottom of the image respectively. We emphasize that the latter representations together constitute the most informative segmentations of (a). Superimposing them (Fig.~\ref{fig:Segmentation}(e)) indeed leads to a good approximation of the original image with only three color levels \{0,127,255\}. The MSR method thus seems to account well for the diversity of content of natural images, inaccessible through classical power spectrum analysis.

\subsection{On the gradient magnitude $|\nabla h|$}

\begin{figure} [t!]
    \centering
    \includegraphics[width = \linewidth]{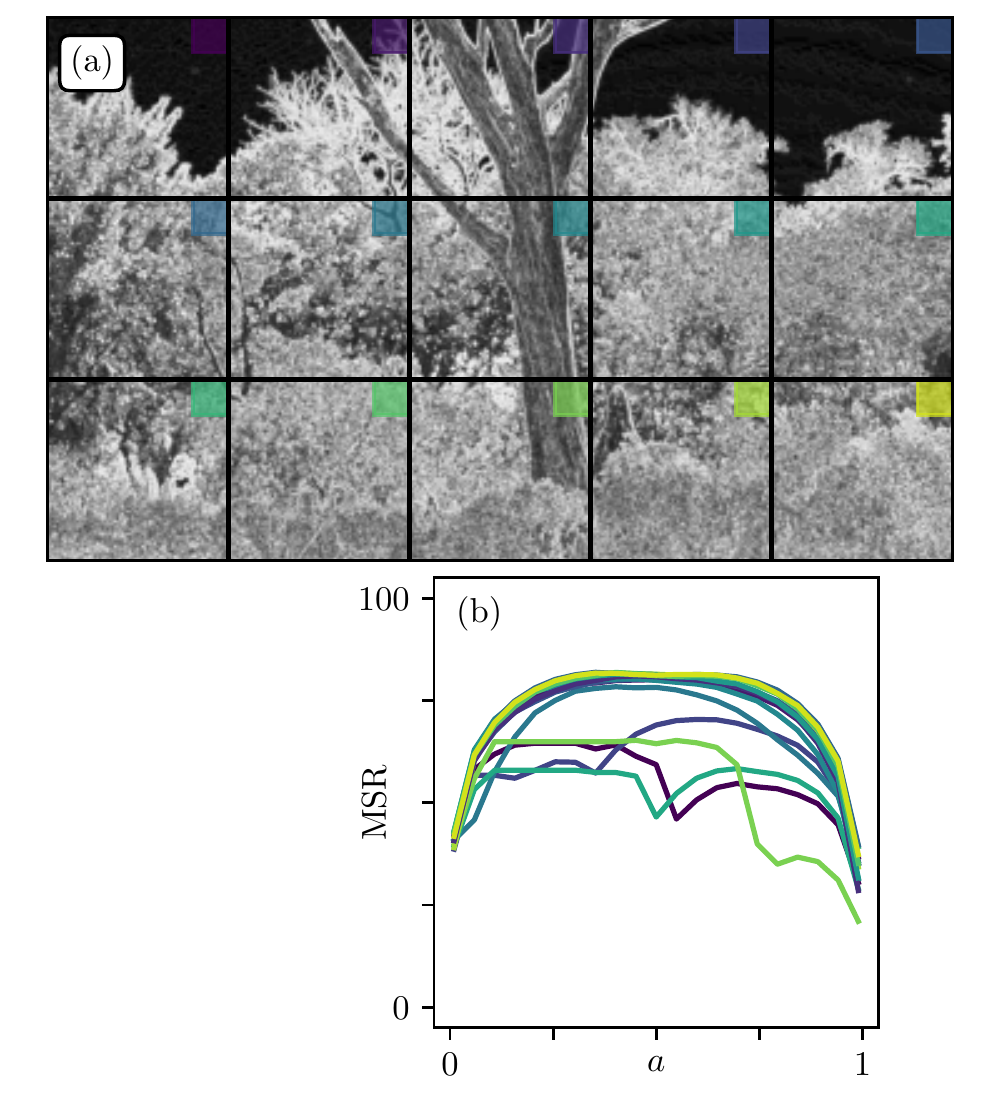}
    \caption{(a) Gradient Magnitude field of Fig.~\ref{fig:natural}(a), with $j=0$, divided in $512\times512$ patches. (b) MSR as function of $a$ for the different patches.}
    \label{fig:Gradient1}
\end{figure}

To understand further the  architecture of natural images, we now focus on the gradient magnitude field intended to capture strong spatial irregularities such as contours or borders. In addition, taking the gradient has the advantage of stationarizing the initial field. The gradient analysis is a fundamental block of various image processing procedures, from  classic edge detection \cite{peli_study_1982}, to  supervised \cite{andreux_kymatio_2020} or unsupervised \cite{krizhevsky_imagenet_2017} classification architectures in machine learning. From a more perception-based psychophysical perspective, it has been shown that essential information such as orientations, geometries and positions could be directly inferred from the visual assessment of the gradient field~\cite{keil_gradient_2006,keil_gradient_2007,kilpelainen_luminance_2018}.
We compute the gradients $|\nabla h|$ from wavelet convolutions. This method is now extensively used as shows excellent robustness for signal processing tasks~\cite{morel2022scale,mallat_multifrequency_1989,antoine_image_1993,abry_wavelet_2009,wendt_multifractal_2009}. On has:
\begin{equation}
    |\nabla h|  = |h \ast \boldsymbol \psi_{j} (\boldsymbol r)|^2,
    \label{eq:LoG}
\end{equation}
where $\boldsymbol \psi_{j} := (\psi_{j,x},\psi_{j,y})$ is a wavelet gradient filter of characteristic dyadic size $2^j$. This wavelet consists in mixing gradient and Gaussian windows, the latter being of standard deviation $\sigma_j = 2^j$ pixels. 
The procedure with $j=0$ yields the image in Fig.~\ref{fig:Gradient1}(a). As expected, one obtains a strong signal (bright shades) for fluctuating textures of vegetation or sharp contours like branches, and low values (dark shades) for smooth and uniform regions like the sky.  

\begin{figure}
    \centering
    \includegraphics[width = \linewidth]{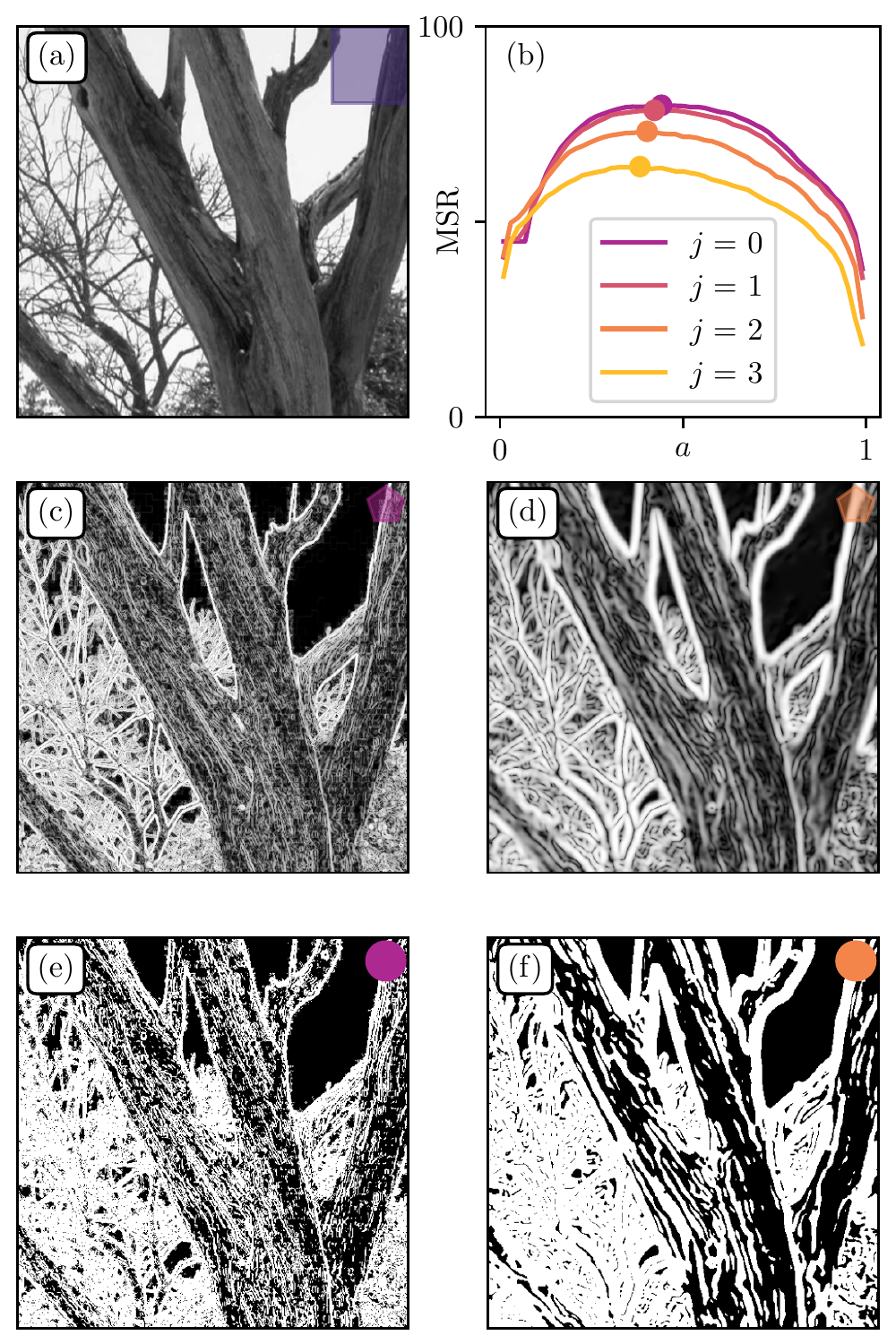}
    \caption{Influence of the Gradient Wavelet size. (a) Original patch from Fig.~\ref{fig:natural}(a). (b) MSR as function of $a$ for gradient wavelets of dyadic size $(2^j), j \in \{0,1,2,3\}$. (c)(d) Gradient magnitudes for $j=0$ and $j=2$ respectively. (e)(f) Segmented gradient magnitudes at critical threshold values $a_c$ for $j=0$ and $j=2$ respectively.}
    \label{fig:Gradient_Thresholding}
\end{figure}

We then conduct the MSR analysis on these new patches (Fig.~\ref{fig:Gradient1}(b)), and observe
that most patches give flat MSR curves. This is tantamount to the critical $\text{H}\approx 0$ case with logarithmic correlations described in Section~\ref{sec:1/f}  (see Fig.~\ref{fig:percolation}). 
One may indeed think of natural images as a patchwork of  objects of various sizes; such superposition of patterns is reminiscent of additive cascades processes~\cite{duplantier_log-correlated_2014} that also display logarithmic correlations.

We now explore the effect of changing the wavelet size (see Fig.~\ref{fig:Gradient_Thresholding}). We chose  the top middle patch in Fig.~\ref{fig:Gradient1}(a) as it contains large objects and small scale details. As one can see in Figs.~\ref{fig:Gradient_Thresholding}(c) and (d), increasing $j$ has the effect of coarse-graining small fluctuations to only leave larger ones. This translates into smaller Relevance at low compression, which in turn reduces the overall MSR (Fig.~\ref{fig:Gradient_Thresholding}(b)). Finally, the segmented gradient fields at critical threshold values (Figs.~\ref{fig:Gradient_Thresholding}(e) and (f)) remain visually  close to initial fields (Figs.~\ref{fig:Gradient_Thresholding}(c) and (d)). This is indeed expected as gradient magnitudes already show a large proportion of black and white pixels at the contours of physical objects.

\section{Application to image processing \label{sec:ImageProcessing}}
Here we illustrate the potential of MSR in the context of common digital image processing tasks, namely color mapping and denoising. 

\subsection{Color mapping}

Consider the color mapping problem consisting in projecting  pixel values onto a reduced palette. For the sake of simplicity, let us consider the case of an initial grayscale palette projected on binary values $\{0,255\}$ (B\&W).
We implement a stochastic mapping procedure using the  Boltzmann distribution $\mathbb P(c|h_{ij})\propto e^{-(h_{ij} - c)^2/T}$, where $h_{ij}$ is the original color of pixel with coordinates $ij$,   $c\in\{0,255\}$ the color in the reduced palette~\footnote{This probability density is obtained from the maximal entropy distribution related to the minimization of the Mean-Squared Error (MSE) between the original and mapped images.}, and $T$  a temperature parameter, see~\cite{lakhal2023new}.  $T=0$  corresponds to the choice of the closest color in the reduced palette, while $T \to \infty$ leads to uniform  noise.

\begin{figure}[t!]
    \centering
    \includegraphics[width = \linewidth]{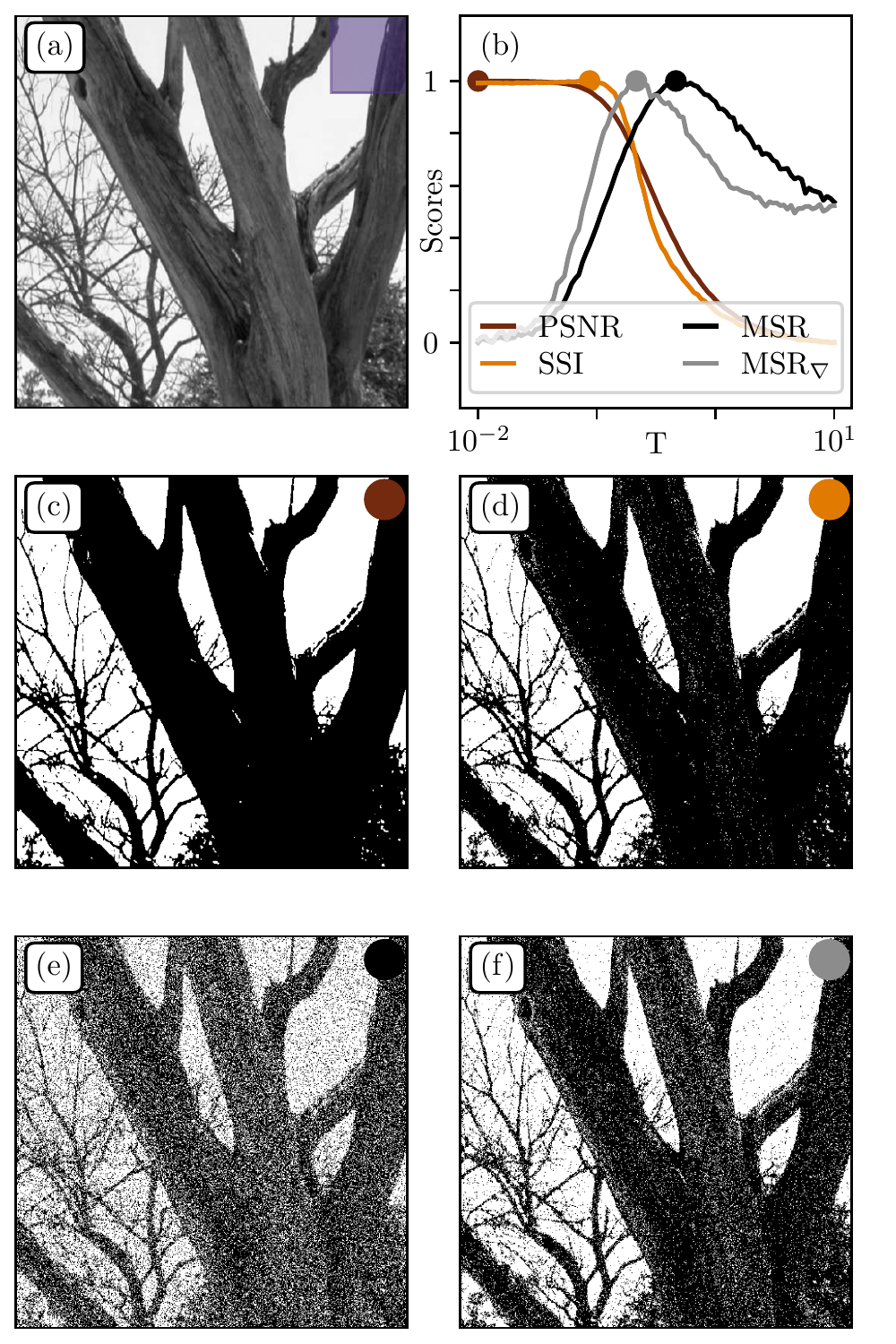}
    \caption{Color mapping. (a) Original patch from Fig.~\ref{fig:natural}(a). (b) Rescaled scores as function of temperature for different performance measures: Peak Signal-to-Noise Ratio (PSNR), Structural Similarity Index (SSI), direct Multiscale Relevance (MSR), and MSR over the gradient field MSR$_\nabla$. (c),(d),(e), (f) Color mapping at optimal temperatures $T^*$ for PSNR, SSI, MSR and MSR$_\nabla$ respectively.}
    \label{fig:ColorReduction}
\end{figure}

Optimizing the procedure consists in calibrating $T$ to maximize some advanced similarity measure between the original and reduced images, in the hope that it will capture more interesting properties than a simple pixel-pixel Euclidian distance minimization. Here we propose an alternative approach consisting in maximizing an information measure, the MSR, and  compare it to classical metrics, namely the Peak Signal-to-Noise Ratio (PSNR)~\cite{korhonen2012peak} and the Structural Similarity Index (SSI)~\cite{wang_image_2004}.
PSNR is directly related to the Mean Squarred Error (MSE) between original and mapped images through PSNR~$=10\log _{10}\left({\Delta^{2}}/{\text{MSE}}\right)$ where $\Delta$ is the range of the signal, that is $255$ for typical grayscale encoding. SSI is based on the comparison of patches between two images and takes into account properties such as luminance and contrast. Both are widely used in the digital image processing community. 

Figure~\ref{fig:ColorReduction}(a) displays the original patch  extracted from Fig.~\ref{fig:natural}(a). Figure~\ref{fig:ColorReduction}(b) shows the evolution of each metrics with temperature $T$. One sees that the PSNR between the original and mapped images is maximized at $T=0$. This is not surprising as the PSNR is monotonously related to the MSE by definition. The corresponding mapping in Fig.~\ref{fig:ColorReduction}(c) appears too sharp and contrasted, clearly separating vegetation from sky while introducing thresholding artifacts. Optimization of the SSI yields a non-zero yet small temperature $T=0.1$, barely improving the resulting image (see Fig.~\ref{fig:ColorReduction}(d)). We then compute the MSR for both direct and gradient fields. The maximization of MSR($T$) leads to the image shown in Fig.~\ref{fig:ColorReduction}(e), which contains more faithful visual features and a decent similarity to the original image at large scales, at the cost of artificial small scale features. Finally, the maximization of the gradient magnitude MSR~\footnote{Note that to compute the gradient magnitude MSR, one has to segment the grayscale images obtained from the gradient procedure, and  average over $a$.}, shown in Fig.~\ref{fig:ColorReduction}(f),  seems like a good compromise between (c),(d) and (e) as it also displays medium scale features (tree trunk details) without blurring finer ones (small branches). 

Hence, for strong color reduction, a Multiscale Relevance approach can bring better visuals than classical metrics such as the Structural Similarity Index which, in addition, requires an \textit{a priori} semantic knowledge of the original image. Note that the analysis could be extended to more elaborate color mapping procedure such as error diffusion \cite{Jarvis_1976,Floyd:1976:AAS} or Monte-Carlo based algorithms \cite{puzicha_spatial_2000}.

\subsection{Denoising with Rudin-Osher-Fatemi algorithm}

We now focus on a denoising procedure which consists in correcting unwanted noise caused by signal processing or camera artefacts. To tackle this problem, a classic algorithm is the Rudin-Osher-Fatemi (ROF) \cite{rudin_nonlinear_1992} which minimizes the following functional:
\begin{equation}
    \mathcal{L}[f] =  \lambda \|\nabla f \|_2^2 +  \| h - f\|_2^2\,,
    \label{eq:ROF}
\end{equation}
where $h$ is the original noisy image, $f$ the target denoised image and $\lambda$ a \textit{regularization/penalty} term preventing gradient explosion and allowing for smooth solutions. The free parameter $\lambda$ is generally chosen by the operator through visual assessment.

\begin{figure}[t!]
    \centering
    \includegraphics[width = \linewidth]{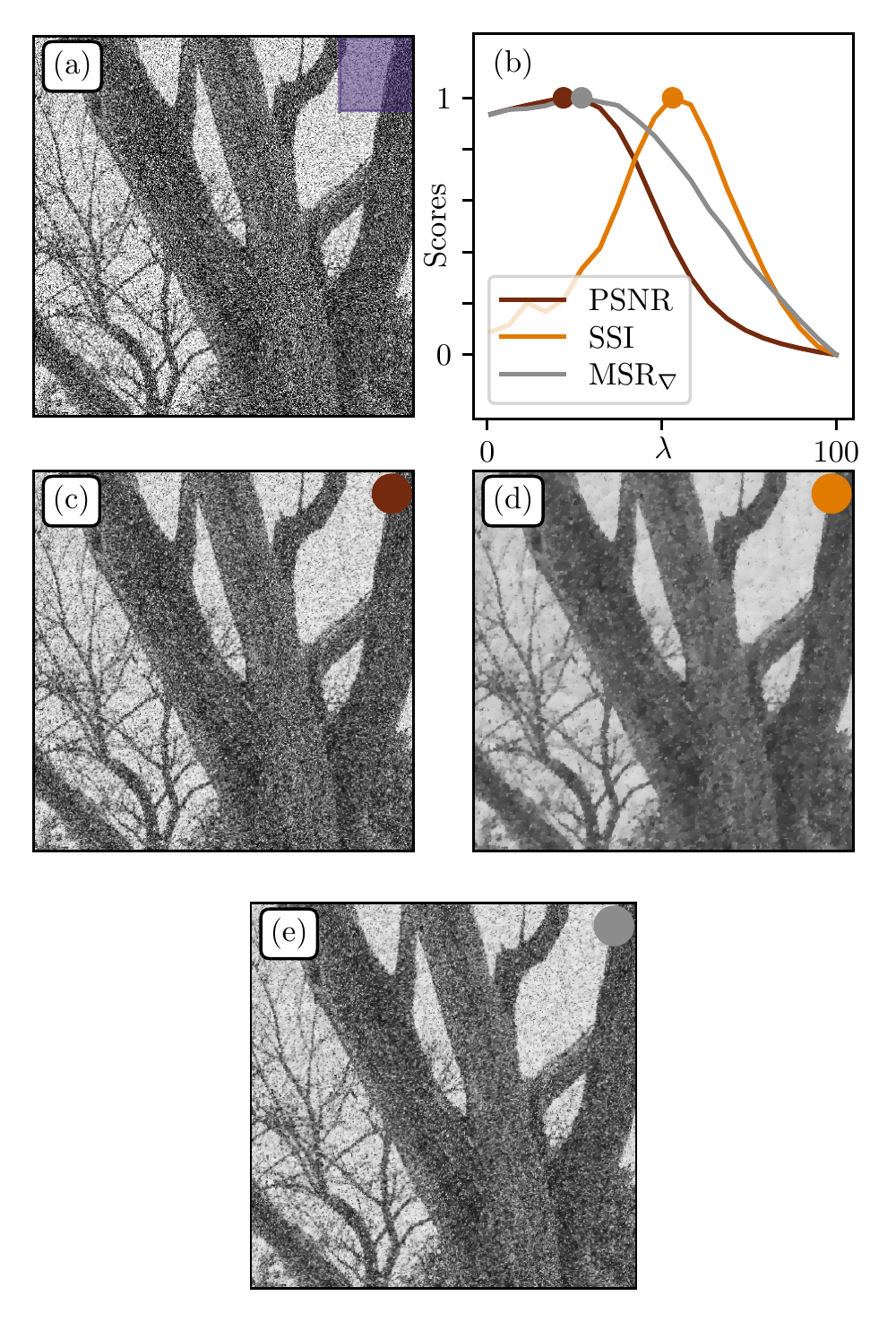}
    \caption{Denoising. (a) Noisy patch obtained from adding a Gaussian noise ($\sigma =100$) to the same patch from Fig.~\ref{fig:ColorReduction}(a). Rescaled scores as function of $\lambda$ for different performance measures: Peak Signal-to-Noise Ratio (PSNR), Structural Similarity Index (SSI) and MSR over the gradient field MSR$_\nabla$. (c),(d), (e) Denoising at optimal regularization parameter $\lambda^*$ for PSNR, SSI and MSR$_\nabla$ respectively.}
    \label{fig:ROFSmoothing}
\end{figure}

Here we propose to calibrate such a model  using again the PSNR, SSI and MSR$_\nabla$ metrics. We consider the image in Fig.~\ref{fig:ROFSmoothing}(a)  obtained by adding a Gaussian white noise to the patch in Fig.~\ref{fig:natural}(a). We intentionally choose a high noise value to make the denoising procedure difficult, such that some details from the original image may never be recovered. Our goal is to seek the optimal $\lambda^*$ leading to the best visual. The scores obtained for each method as function of $\lambda$ are displayed in Fig.~\ref{fig:ROFSmoothing}(b). Optimally denoised images using PSNR, SSI and MSR$_\nabla$ are shown in Fig.~\ref{fig:ROFSmoothing}(c), (d) and (e) respectively. With PSNR, one is left with a rather high level of noise, while details on the trunk surface or in the branches are conserved. In contrast, SSI removes a significant part of the noise, but at the cost of blurring small scale details. Although less obvious than for the color mapping procedure, optimal denoising with MSR$_\nabla$ seems like a good compromise between a too noisy PSNR image and an overly smoothed SSI image.

\section{Conclusion}

Let us summarize what we have achieved. We first introduced the Resolution/Relevance framework through a simple illustrative example. We showed how such formalism can be applied to image analysis. With the aim of investigating the framework in a controlled environment,  we started by studying random textures. We then defined the Multiscale Relevance (MSR) which measures the entropy contribution at all compression scales, and obtained statistical features reminiscent of the correlated percolation problem. In particular, we highlighted the existence of a critical roughness parameter $\text{H}_c \approx 0$, corresponding to logarithmic correlations, and discussed optimal segmentation.  We then extended the analysis to natural images and drew a successful comparison with random textures;  we observed  strong similarities with critical random Gaussian fields. Looking at gradient magnitude fields revealed an even stronger similarity to roughness criticality. Finally, we confronted the MSR procedure to classical signal processing measures in the context of simple image processing tasks: color mapping and denoising. We obtained interesting results thereby demonstrating the potential of the agnostic MSR approach for image processing.

This last section would benefit from an extension to more elaborate image processing techniques, beyond the scope of the present paper. Future research should also focus on analytically tractable developments of Relevance and Resolution in simple cases, e.g. Gaussian white noise with well chosen cascading processes. Also note that we considered a straightforward compression procedure on the direct space but equivalent representations, for example Discrete Cosine~\cite{wallace_jpeg_1992} or Wavelet harmonics~\cite{skodras_jpeg_2001}, could be used to define the reduced sample $\mathcal S$. Finally, we have seen that the MSR is able to capture the most relevant segmentation values, which may be used as a pre-processing method for learning frameworks.

\section{Acknowledgments}

We deeply thank Jean-Philippe Bouchaud for fruitful discussions.
This research was conducted within the Econophysics \& Complex Systems Research Chair, under the aegis of the Fondation du Risque, the Fondation de l'Ecole polytechnique, the Ecole polytechnique and Capital Fund Management. 
S.L. would like to thank the Quantitative Life Science group for his stay at the Abdus Salam International Centre for Theoretical Physics (ICTP) in Trieste, Italy.
\bigskip

\appendix

\section{Generation of $1/f^{\alpha}$ Gaussian fields \label{appendix:1/f}}

The $1/f^\alpha$ textures used in Sec.~\ref{appendix:1/f} are generated through a filtering procedure. 
 Consider a field $\phi(\boldsymbol r)$ of given autocorrelation function $C(\boldsymbol r)$. Using the convolution theorem one has 
$
\tilde C(\boldsymbol q) =  |\tilde \phi(\boldsymbol q) |^2 
$
in Fourier space with $|\boldsymbol q| = f$. A natural way to generate a random Gaussian field $\phi$ with prescribed correlations is:
\begin{equation}\phi =  \mathcal{F}^{-1}\left(\sqrt{|\tilde C(\boldsymbol q)|}\tilde \eta(\boldsymbol q)\right),
\end{equation}
where $\eta(\boldsymbol r)$ is a  Gaussian white noise.
Note that, as a result, the textures display periodic boundary conditions.
Also note that the $\text{H}=0$  (equivalently $\alpha = 1$) case corresponds to logarithmic spatial correlations of the form (see e.g.~\cite{peskin_introduction_1995}):
\begin{equation}
    C(\boldsymbol r) = -\lambda\left(\log\frac{r}{\xi} + \log 2 - \gamma\right),
\end{equation}
where $r=|\boldsymbol r|$, $\xi^{-1}$ is a regularizing constant that can be interpreted as a low frequency cutoff, and $\gamma$ is the Euler constant.

\bibliographystyle{apsrev4-1}
\nocite{apsrev41Control}
\setlength{\bibsep}{1.8pt plus 0.3ex}
\bibliography{lakhal2023}

\end{document}